\documentclass[twocolumn,superscriptaddress,english,prl,showpacs,longbibliography]{revtex4-2}
\usepackage[colorlinks=true,urlcolor=blue,citecolor=blue,linkcolor=blue]{hyperref} 
\usepackage{amsmath}
\usepackage{graphicx}
\usepackage{bm}
\usepackage{hyperref}
\usepackage{color}
\usepackage{amssymb}
\usepackage{graphicx}
\usepackage{subfigure} 
\usepackage{color}
\usepackage{mathrsfs}
\usepackage{float}
\usepackage{indentfirst}
\usepackage{txfonts}
\usepackage{algpseudocode}  
\usepackage{balance}
\newcommand{\s}{\mathbf {s}}
\newcommand{\x}{\mathbf {x}}
\newcommand{\vv}{\mathbf {v}}
\newcommand{\hh}{\mathbf {h}}
\newcommand{\wt}{\widetilde}
\newcommand{\wtp}{\widetilde P}
\newcommand{\wttp}{\widetilde {\mathcal P}}

\newcommand{\ii}{\mathbf {\mathcal I}}
\newcommand{\one}{ \left(\begin{array}{c}1\\1\end{array}\right)}
\newcommand{\dmax}{D_{\mathrm{max}}}
\usepackage[keeplastbox]{flushend}
\begin{document}

\preprint{APS/123-QED}

\title{Boltzmann machines as two-dimensional tensor networks}
\author{Sujie Li}
\affiliation{
 CAS Key Laboratory for Theoretical Physics, Institute of Theoretical Physics, Chinese Academy of Sciences, Beijing 100190, China
}
\affiliation{
 School of Physical Sciences, University of Chinese Academy of Sciences, Beijing 100049, China
}
\author{Feng Pan}
\affiliation{
 CAS Key Laboratory for Theoretical Physics, Institute of Theoretical Physics, Chinese Academy of Sciences, Beijing 100190, China
}
\affiliation{
 School of Physical Sciences, University of Chinese Academy of Sciences, Beijing 100049, China
}
\author{Pengfei Zhou}
\affiliation{
 CAS Key Laboratory for Theoretical Physics, Institute of Theoretical Physics, Chinese Academy of Sciences, Beijing 100190, China
}
\affiliation{
 School of Physical Sciences, University of Chinese Academy of Sciences, Beijing 100049, China
}

\author{Pan Zhang}
\email{panzhang@itp.ac.cn}
\affiliation{
 CAS Key Laboratory for Theoretical Physics, Institute of Theoretical Physics, Chinese Academy of Sciences, Beijing 100190, China
}
\affiliation{School of Fundamental Physics and Mathematical Sciences, Hangzhou Institute for Advanced Study, UCAS, Hangzhou 310024, China}
\affiliation{International Centre for Theoretical Physics Asia-Pacific, Beijing/Hangzhou, China}

\begin{abstract}
Restricted Boltzmann machines (RBM) and deep Boltzmann machines (DBM)
are important models in machine learning, and recently found numerous applications in quantum many-body physics. We show that there are fundamental connections between them and tensor networks. In particular, we demonstrate that any RBM and DBM can be exactly represented as a two-dimensional tensor network. This representation gives an understanding of the expressive power of RBM and DBM using entanglement structures of the tensor networks, also provides an efficient tensor network contraction algorithm for the computing partition function of RBM and DBM. Using numerical experiments, we demonstrate that the proposed algorithm is much more accurate than the state-of-the-art machine learning methods in estimating the partition function of restricted Boltzmann machines and deep Boltzmann machines, and have potential applications in training deep Boltzmann machines for general machine learning tasks.
\end{abstract}

\maketitle
Inspired by statistical physics, the Boltzmann machine was invented more than $30$ years ago~\cite{hinton1983optimal}, as a powerful model for representing a joint distribution of high dimensional data using the Boltzmann distribution, 
\begin{equation}\label{eq:Boltzmann}
   P(\vv) = \frac{1}{Z}\wtp = \frac{1}{Z}\sum_{\hh}e^{-\beta E(\vv,\hh)},
\end{equation}
where $Z$ is the partition function, and configuration $\vv\in \{+1,-1\}^n$ denotes a data vector composed of $n$ binary components, representing a collection of data variables e.g. $n$ pixels in a natural image, and $\hh\in\{+1,-1\}^m$ denotes a configuration of $m$ hidden variables. Usually, the inverse temperature $\beta$ is set to $1$, and the energy function $E$ contains all the model parameters.

\begin{figure*}[htb]
    \centering
    \includegraphics[width=1.8\columnwidth]{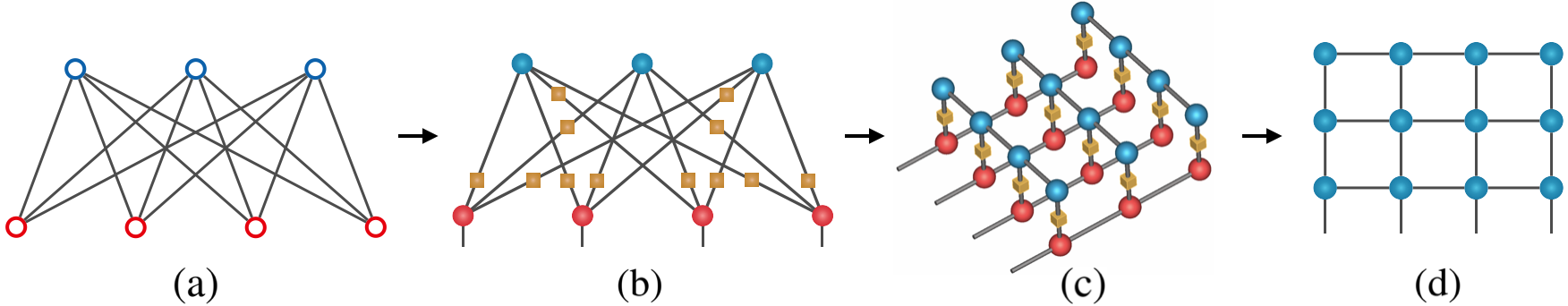}
\caption{Mapping an RBM to a two-dimensional tensor network. (a) RBM with $4$ visible neurons (red circles) and $3$ hidden neurons (blue circles). (b) A tensor network representation of RBM, the red nodes and blue nodes are copy tensors corresponding to visible variables and hidden variables respectively, and the brown squares denote Boltzmann matrices between visible nodes and hidden nodes. (c) The three-dimensional diagram after converting each copy tensor to an MPS. (d) The final two-dimensional tensor network representation of the RBM.}
\label{fig:contraction}
\end{figure*}

A variant of the Boltzmann machine, the restricted Boltzmann machine (RBM) was invented later on, with the connectivity graph of neurons restricted to a bipartite graph where there are only couplings between visible and hidden neurons (as shown in Fig.~\ref{fig:contraction} (a)). The energy function of RBM is defined as
\begin{align}
    E(\vv,\hh)=-\sum_{(ij)}J_{ij}v_ih_j-\sum_{i}\theta_iv_i-\sum_{j}\theta_jh_j,
\end{align}
where $J_{ij}$ denotes coupling between visible neuron $i$ and hidden neuron $j$, $\theta_i$ and $\theta_j$ denote external field of neuron $i$ and $j$ respectively.
Another variant of Boltzmann machines is the deep Boltzmann machine (DBM)~\cite{hinton2006reducing,salakhutdinov2012efficient,salakhutdinov2008learning}, where more than $2$ layers of hidden neurons are allowed.

RBM and DBM play important roles in modern machine learning~\cite{Goodfellow2016Deep}. For example, using DBM as an efficient way for pre-training the deep neural networks~\cite{hinton2006reducing} effectively initiated the fruitful research area of ``deep learning'', also finding applications in classification problems~\cite{larochelle2008classification}, recommendation systems~\cite{salakhutdinov2007restricted}, representation learning in speech and sound systems~\cite{jaitly2011learning}, time-series analysis~\cite{kuremoto2014time}, etc.
RBM and DBM also found numerous applications in physics recently~\cite{ carleo2019machine,melko2019restricted}. They have been used for accelerating the Monte Carlo algorithm~\cite{huang2017accelerated}, as a variational ansatz in variational Monte Carlo methods\cite{Carleo2017}, as a good representation for quantum states in many-body systems~\cite{carleo2018constructing,lu2019efficient,gao2017efficient,nomura2017restricted,deng2017quantum}, and for quantum state tomography~\cite{torlai2018neural}. 

There are two properties of RBM that are widely appreciated. The first one is the representation power. It has been proved in~\cite{le2008representational} that RBMs are universal approximators of discrete probability distributions, meaning that with enough hidden variables they can represent any probability distribution. For representing quantum states, in~\cite{deng2017quantum} it has been shown that RBM exhibits volume-law entanglements. This originates from the long-range interactions between hidden neurons and visible neurons, which can induce complex correlation structures among visible neurons, particularly in the context of quantum state representations, where standard variational ansatz for quantum states usually involve only local and short-range interactions. The second recognized property of RBM is that the bipartite topology of the model allows an efficient block update in Gibbs sampling, and for efficient estimate of expectations and correlations, based on which scalable inference and learning algorithms can be implemented.

However, there are also some limitations of RBM associated with the properties. The first one is about the efficiency of the representation power. Although in principle RBM can represent any distribution with an unlimited number of hidden neurons, it is not clear how to quantitatively characterize the representation power of RBM with a given number of hidden neurons. This gives rise to serious doubts on the efficiency of the representation power of RBM. 
This limitation motivates the deep Boltzmann machine, which contains multiple layers of hidden neurons. DBM uses RBM as a building block and packs them one by one to form a deep model.
Although there exist mathematical proofs that adding more hidden layers strictly improves the representation power~\cite{le2008representational}, and the DBM can efficiently represent some physical ground states which can not be represented by the RBM~\cite{gao2017efficient}, there is still a lack of good understandings of the representation power of DBM and the detailed difference between the expressive power of DBM and RBM in general situations.

Second, although there exist efficient sampling methods with block updates~\cite{hinton2002training}, it is difficult to compute the exact probability of configurations, because there is no efficient method for computing the normalization, i.e. the partition function of the Boltzmann distribution. This is considered as a limitation of RBM in unsupervised machine learning, because an important indicator to evaluate how good the model is in representing the data distribution is the negative log-likelihood $\mathcal L = -\sum_{\x\in\mathrm{data}}\log P(\x)$,
which requires an access to the explicit value of probability $P(\x)$.
In the machine learning community, researchers rely on MCMC based methods, such as the annealed importance sampling (AIS)~\cite{salakhutdinov2010learning}, to estimate the partition function and the negative log-likelihood. 
For DBM, in addition to the difficulties in computing the partition function, it also suffers from slow sampling and training due to the lack of efficient block update strategies.

In this article we explore how to understand RBM and DBM from the viewpoint of tensor networks, targeting two limitations discussed above, the characterization of the representation power and the computation of the partition function. We show that any RBM and DBM can be mapped exactly to two-dimensional tensor networks, thus the representation power of them can be studied using entanglement structures of the tensor network. Moreover, the two-dimensional tensor network representation implies that the partition function of RBM and DBM can be computed using tensor network contractions methods e.g. tensor network renormalization group approaches. Using numerical experiments we show that the tensor network contraction method is more accurate than the existing method adopted in machine learning for the computing partition function of RBM and DBM.

\paragraph{Mapping RBM and DBM to two-dimensional tensor networks---}
Any probability distribution $P(\s)$ and its un-normalized version $\wt P(\s)$ (as in Eq.~\eqref{eq:Boltzmann}) of discrete variables $\s$ can be regarded as a giant tensor with non-negative entries living in a linear space with dimension $2^n$, where $n$ is the number of visible variables, and $2$ comes from our assumption that (without loss of generality) every variable takes $2$ states. The difference of representations for $P$ and $\wt P$ is that all elements of the tensor $\mathcal P$ corresponding to $P(\s)$ sum to $1$ by definition of the normalized probability, while summation of all elements of tensor $\wttp$ corresponding to $\wtp(\s)$ equals the partition function $Z$. In this sense, the computation of partition function can be mapped to the problem of computing the $\ell_1$ norm of $\wttp$, which is the inner product of $\wttp$ and all $1$ vector $\mathbf 1_{2^n}=\underbrace{1,1,1,\cdots,1}_{2^n}$. 
At the first glance, the computation looks intractable because an exponential large space is involved. However if we could find a tensor network representation of $\wttp$ with a proper structure, we might be able to compute the inner product efficiently.
\begin{figure}[h]
\centering
\subfigure[]{
\label{fig:mps1}
\includegraphics[width=0.31\columnwidth]{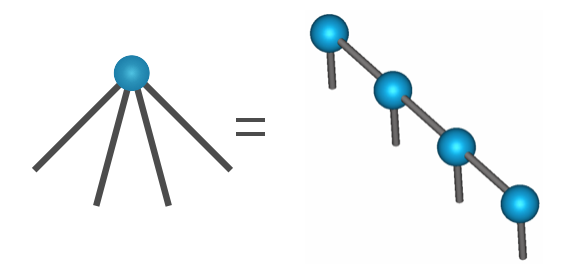}}
\subfigure[]{
\label{fig:mps2}
\includegraphics[width=0.31\columnwidth]{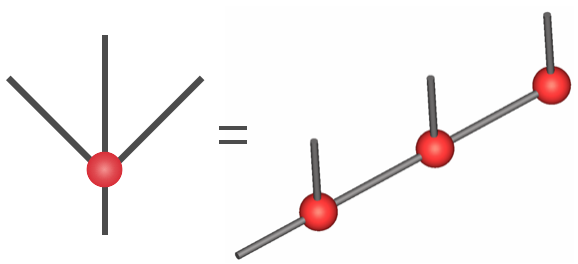}}
\subfigure[]{
\label{fig:element}
\includegraphics[width=0.31\columnwidth]{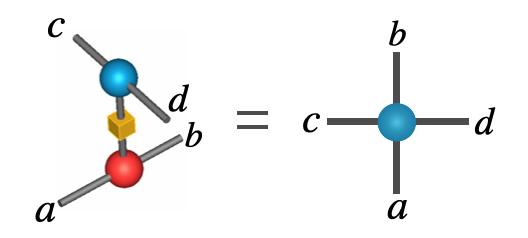}}
\caption{(a) and (b) are schematic diagrams of converting a copy tensor to an MPS. (a) corresponds to an hidden variable, while (b) corresponds to a visible variable which has a open index for representing  a variable in $\wttp$. (c) Converting the three-dimensional structure to a single tensor.}
\label{fig:mps}
\end{figure}

\begin{figure*}[htb]
\centering
\includegraphics[width=1.8\columnwidth]{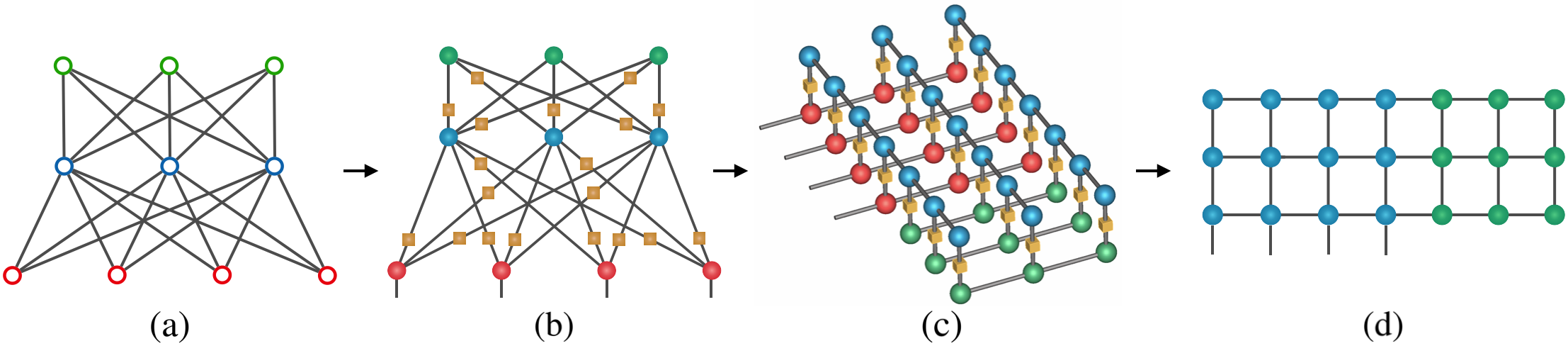}
\caption{Mapping a DBM with two hidden layers to a two-dimensional tensor network, the symbols have the same meaning with Fig.~\ref{fig:contraction}}
\label{fig:dbm1_all}
\end{figure*}

In the case of RBM, we can express $\wttp$ as a two-dimensional tensor network. The detailed procedure is illustrated in Fig.~\ref{fig:contraction}. The original RBM is shown in Fig.~\ref{fig:contraction}(a), where the red circles denote the visible variables and the blue circles denote the hidden variables, and the black lines denote couplings $J_{ij}$ connecting two kinds of variables. The first step of the conversion is to obtain un-normalized tensor $\wttp$, by introducing copy tensor $\ii^{(i)}$ for variable $i$ (on circles), and Boltzmann matrices $\mathbf B_{ij}$ for coupling $J_{ij}$ (on black lines), as shown in Fig.~\ref{fig:contraction}(b). The copy tensor $\ii^{(i)}$ is a diagonal tensor with an order equal to the number of neighbors of node $i$, and with $1$ on the diagonal entries and zero on the other entries.
The Boltzmann matrices encode the interactions between a pair of variables \begin{align}
\mathbf{B}_{ij}=\left(               
  \begin{array}{cc}   
    e^{J_{ij}} &e^{-J_{ij}} \\  
    e^{- J_{ij}}& e^{ J_{ij}} \\  
  \end{array}
\right). \label{B1}  
\end{align}
 The blue copy tensors have no open indices because they have been traced out.
We can see that this step maps the RBM to a tensor network with the same topology, which is still difficult to deal with. 

Next, we use the mathematical properties of the copy tensors to manipulate the topology. First we notice that a $l$-way copy tensor $\ii_l$ can be converted to an matrix product state with length $l$, i.e. $\ii_l=\underbrace{\ii_2\times \ii_3\times\cdots\times \ii_3\times \ii_2}_{l}$,
with $\times$ symbol representing tensor contraction. 
In Appendices, we gave proof for the conversion. Another way to understand the conversion is that the copy tensor is a special case of the canonical polyadic (CP) tensor format~\cite{chi2012tensors}, with CP rank equals $2$. It is known in mathematics that there exists an exact conversion from the CP tensor with rank $2$ to a matrix product state with bond dimension $2$~\cite{oseledets2011tensor}. 
The diagram representation of the conversion is illustrated in Fig.~\ref{fig:mps}, where both blue copy tensors corresponding to the visible variables, and the red copy tensors corresponding to the hidden variables, are converted to MPSs in the same way, leaving the brown Boltzmann matrices $\mathcal B$ connecting blue and red MPSs. This leads to a quasi three-dimensional tensor network representation of $\wttp$ as shown in Fig.~\ref{fig:contraction}(c).
Then by contracting the brown Boltzmann matrix with the red and blue tensors (as shown in Fig.~\ref{fig:mps2}), we finally arrive at Fig.~\ref{fig:contraction}(d), the two-dimensional tensor network representation $\wttp$ of the un-normalized probability tensor of RBM. Using this method, the RBM with $n$ visible neurons and $m$ hidden neurons can be exactly mapped to a two-dimensional tensor network with size $n\times m$.
For simplicity, we did not consider external fields. The conversion procedure is the same when external fields (or bias) present, the details together with precise mathematical formulations of the conversions are shown in Appendices.

In the literature, there are two alternative tensor network representations of RBM. In~\cite{Chen2018Equivalence}, it has been shown that an RBM can be mapped to an MPS, possibly with exponentially large bond dimensions. 
The MPS representation of RBM can be obtained straightforwardly from our representation, by further contracting the two-dimensional tensor network from top to bottom, resulting in an MPS with possibly exponential bond dimensions. In this sense, our representation is more fine-grained than the MPS representation. 
It has been shown in~\cite{glasser2018neural} that a short-range RBM corresponds to the Entangled Plaquette State (EPS) while a fully connected RBM corresponds to a string bond state (SBS). 
Notice that the EPS and SBS representations focus on the expressive power of RBM, while our representation also provides an accurate contraction algorithm for computing the partition function, and generalizes to its deep version, DBM, as we will introduce here.

\begin{figure}[h]
\flushleft 
\subfigure[]{
\label{fig:rbm1}
\includegraphics[width=0.3\columnwidth]{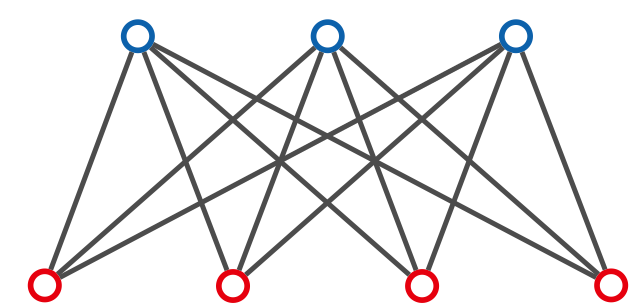}
}
\subfigure[]{
\label{fig:dbm10}
\includegraphics[width=0.3\columnwidth]{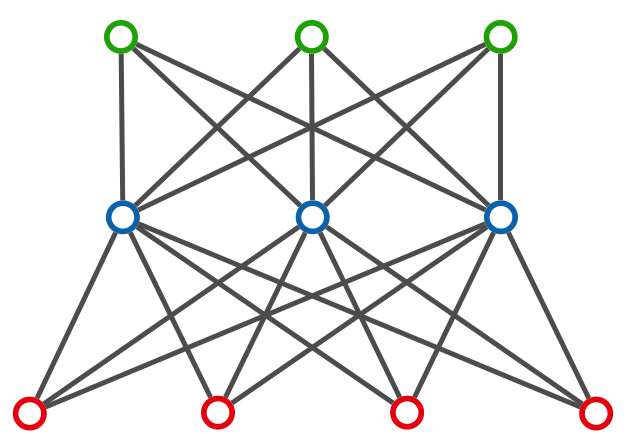}
}
\subfigure[]{
\label{fig:dbm20}
\includegraphics[width=0.3\columnwidth]{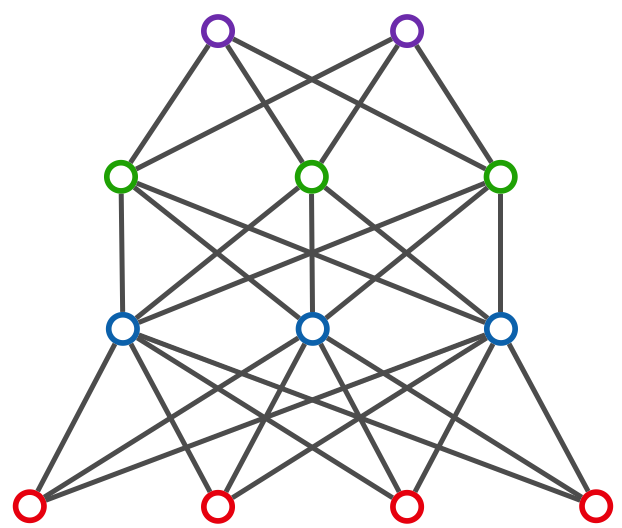}}
\subfigure[]{
\label{fig:tn}
\includegraphics[width=0.3\columnwidth]{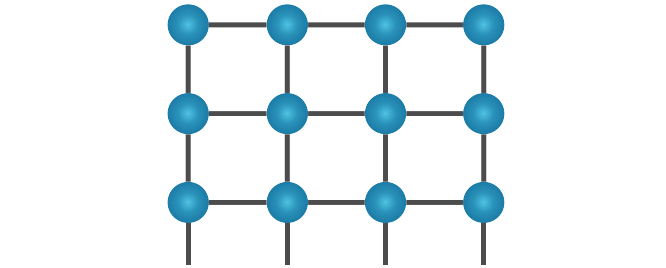}}
\subfigure[]{
\label{fig:dbm1_tn}
\includegraphics[width=0.3\columnwidth]{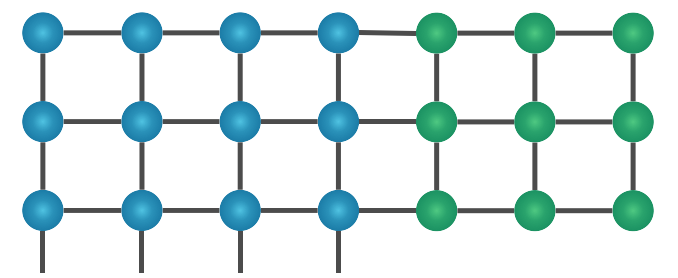}
}
\subfigure[]{
\label{fig:dbm2_tn}
\includegraphics[width=0.3\columnwidth]{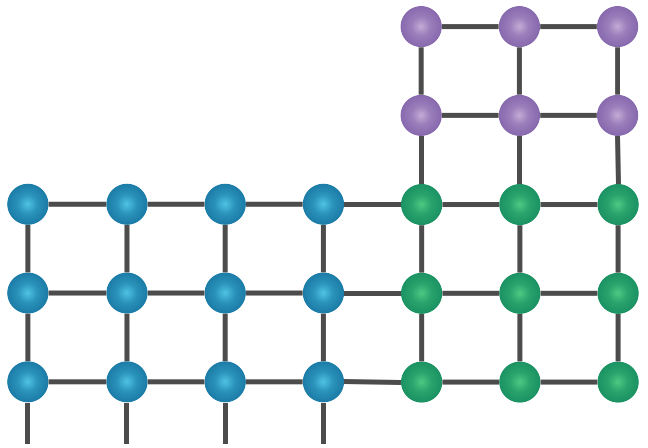}}
\quad 
\caption{(a) An RBM with $4$ visible neurons and $3$ hidden neurons. (b) A DBM with two hidden layers (blue and green). (c) A DBM with $3$ hidden layers. (d),(e),(f) are the two-dimensional tensor network representations of (a), (b) and (c) respectively.}
\label{fig:dbm}
\end{figure}

When considering a DBM with more than one hidden layer (as shown in Fig.~\ref{fig:dbm}), we use the same method to first represent the DBM as a tensor network composed of copy tensors and Boltzmann matrices with the same topology (as shown in Fig.~\ref{fig:dbm1_all}(b)), then generalize the mapping method in RBM to DBM, by converting every copy tensor to an MPS (Fig.~\ref{fig:dbm1_all}(c)), forming a tensor network as shown in Fig.~\ref{fig:dbm1_all}(d). The resulting tensor networks for RBM, DBM with more hidden layers are summarized in Fig.~\ref{fig:dbm}. More examples of conversion can be found in the appendices. 
A feature we can extract from the representations (in Fig.~\ref{fig:dbm}) is that no matter how many hidden layers a DBM has, the tensor network corresponding to the DBM is always in two dimensions. By inspecting the entanglement structure of the tensor network, one can find that the entanglement entropy follows area law in the two-dimensional $n\times m$ lattice, where $n$ is the number of visible neurons and $m$ is the number of hidden neurons in the {first layer}. This gives rise to a surprising conclusion that the upper limit of the expressive power of DBM depends only on the \textit{number of neurons in the first hidden layer}, the same as RBM. Here we note that in practice the representation power of DBM could be much better than RBM because the converted tensor network is not a general two-dimensional tensor network as all the tensors in the converted tensor network are sparse, so the upper limit is actually not reachable for RBM and adding more layers in DBM may help in reducing the gap to the limit. In other words, DBM may have better parameter efficiency than RBM in representation power. This is consistent with previous studies in the unsupervised machine learning~\cite{salakhutdinov2012efficient} which showed that 
as a generative model for standard image datasets, DBM gives better log-likelihood in describing the data and is also consistent with previous studies in the quantum physics~\cite{gao2017efficient} which showed that there exist some quantum states that can be represented by DBM but can not be represented by RBM.

\paragraph{Computing the partition function by contracting the two-dimensional tensor network---}\label{sec:contraction}
Computing the partition function is required in obtaining the data likelihood, and in evaluating the model. However, exactly computing the partition function of RBM and DBM belongs to the class of \#P hard problems, there are no polynomial algorithms for solving it in general. There have been extensive studies in machine learning community~\cite{Goodfellow2016Deep,salakhutdinov2008learning} and statistical physics community~\cite{NIPS2015_5788,Huang2015} on how to approximately estimate the partition function. These include variational methods such as belief propagation, and Thouless-Anderson-Parlmer equations, and Monte Carlo methods such as Wang-Landau method, and Annealed Importance Sampling. 

The two-dimensional tensor network representation provides an alternative method for computing the partition function of RBM and DBM using tensor network contractions. We have shown that the partition function can be computed using the inner product of the tensor $\wttp$ and the all one vector $\mathbf 1_{2^n}$. With the tensor network representation of $\wttp$ as illustrated in Fig.~\ref{fig:dbm}. Then one can further factorize the $\mathbf 1_{2^n}$ vector as a tensor product of $n$ small $(1,1)$ vectors, which gives
\begin{equation}
Z =\left\|\wttp\right\|_1  = \wttp \cdot  \mathbf 1_{2^n}^{\top} = \wttp \cdot   \underbrace{ \one\otimes \one\otimes\cdots\otimes \one}_{n} ,
\label{eq:Z}
\end{equation}
where each $\one$ vector is contracted to an open index of the two-dimensional tensor network. This finally converts the computation of partition function to the contraction problem of two-dimensional tensor networks without open indices. There exist many methods for contacting two-dimensional tensor networks~\cite{levin2007tensor,PhysRevB.86.045139,PhysRevLett.91.147902,PhysRevLett.118.110504,2019Contracting} efficiently, in this work we adopt the boundary matrix product states method, which first identifies a boundary of the network, treats it as a matrix product state and other parts of the network as a stack of matrix product operators (MPO). During the contraction process, in each step the MPS on the boundary is contracted to an MPO, resulting to an MPS with larger bound dimensions. If the bond dimension is larger than the pre-determined maximum value $\dmax$, we convert the MPS to the canonical form~\cite{schollwock2011density} and perform singular value decompositions to reduce the bond dimension to $\dmax$. This procedure is similar to the density matrix renormalization group~\cite{white1992density}. We refer to the appendices for detailed descriptions of the contraction algorithm.

To evaluate the performance of the proposed tensor network contraction algorithm in computing partition function of RBM and DBM, we perform numerical experiments on RBM and DBM with random couplings, compute $\ln(Z)$ values for each of them and evaluate the results against the standard annealed importance samplings method (AIS)~\cite{neal2001annealed} which has been widely used in learning and evaluating Boltzmann machines~\cite{salakhutdinov2008learning}. 
The AIS method converts the computation of the partition function to the computation of ratio between the target partition function and the partition function of a known distribution, which is further approximated as the product of a series of ratios between consecutive distributions belonging to a sequence of intermediate distributions~\cite{neal2001annealed}.

\begin{figure}[htb]
    \centering
    \includegraphics[width=\columnwidth]{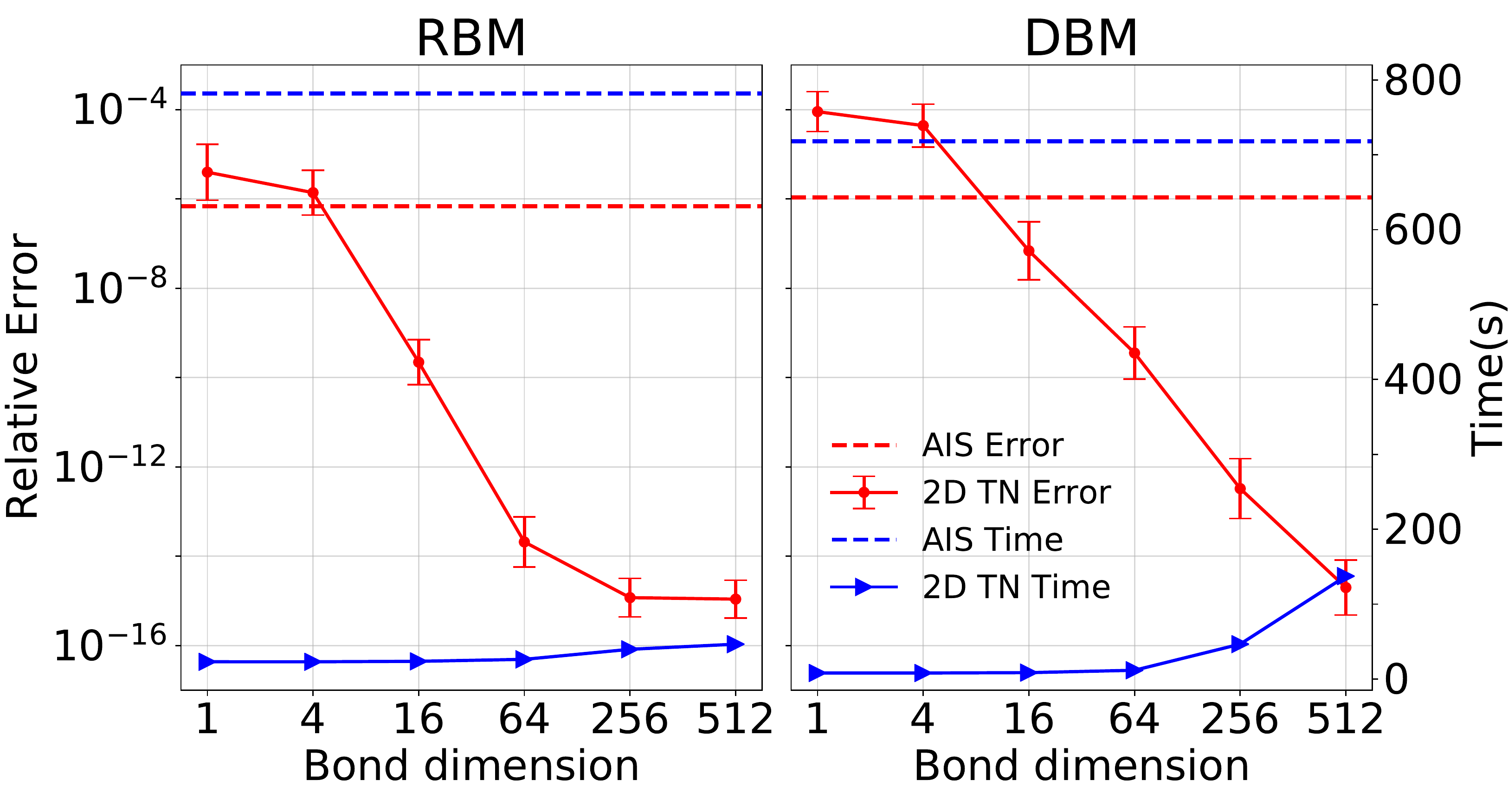}
    \caption{Comparison of the running time and the error on the log partition function per visible variable given by our tensor network contraction algorithm (2D TN) and of AIS algorithm on RBM and DBM with random couplings. Each data point is  averaged over $100$ instances, and the error bar represent one standard derivation. For AIS, we use $100000$ intermediate distributions and $1000$ runs.}
    \label{fig:AIS}
\end{figure}

In the experiments, the RBM instances contain $n=20$ visible variables and $20$ hidden variables, and DBM has one visible layer and two hidden layers, each of which contains $20$ variables.
We choose the size of the RBM and DBM instances in such a way that the exact computation of $\ln(Z)$ is tractable in a reasonable time so that we can evaluate the error of the computations for different methods.
For AIS, we use a large number of intermediate distributions and $1000$ AIS runs, which is the most accurate results we can obtain with AIS. The results are averaged over $100$ random instances where the weights and biases are sampled from a normal distribution with $0$ mean and variance $1/n$ (where $n$ is the number of variables).
The results are shown in Fig.~\ref{fig:AIS}, where we can see that with bond dimensions larger than $4$ in RBM and larger than $10$ in DBM, our method gives a much smaller error than AIS in estimating $ln(Z)$. The figure also shows that for all bond dimensions in our experiments, the tensor network contraction algorithm takes a much shorter running time than AIS which needs time-consuming MCMC samplings.
\paragraph{Discussions---}
We have presented an exact representation of RBM and DBM using two-dimensional tensor networks. This representation gives us a systematic way to characterize the expressive power of RBM and DBM, also provides an efficient algorithm for computing the partition function of RBM and DBM based on tensor network contractions. Apparently, the ability of computing partition function using tensor network contractions immediately gives us the power to compute the negative log-likelihood of training data, which can be set as a loss function in unsupervised machine learning for training model parameters. We will put the study of applications of our method in training RBM and DBM into future work.
We also anticipate that tensor network representations can be generalized to other neural network models, for understanding their representation and providing efficient learning algorithms.

\acknowledgements
P.Z. is supported by Key Research Program of Frontier Sciences, CAS, Grant No. QYZDB-SSW-SYS032, and Project 12047503 and 11975294 of National Natural Science Foundation of China. Part of the computation was carried out at the High-Performance Computational Cluster of ITP, CAS.
\bibliography{main.bib}
\newpage
\clearpage
\onecolumngrid
\appendix  
\section{Converting copy tensors to MPSs}
Here we consider the building block of the mapping from RBM and DBM to two-dimensional tensor networks, the conversion from a copy tensor to an MPS. For simplicity we consider the case when the local dimension of each RBM and DBM variable is $2$. A $l$-way copy tensor $\ii_l$ is a $l$-way tensor with two diagonal entries being $1$ and other entries being zero, written as
\begin{equation}
\ii_l(i_1,i_2,...,i_l) =\left\{ \begin{matrix}1, & \textrm{if } i_1=i_2=...=i_l,\\0, & \textrm{otherwise.} \end{matrix}\right.     = \prod_{a=1}^{l-1}\delta(i_a,i_{a+1}),
\end{equation}
where $\delta(i_a,i_{a+1})$ denotes a delta function.
Suppose we have an MPS $\mathcal A$ which is composed of two identity matrices and $l-2$ copy tensors $\ii_3$, that is,
$ \mathcal A = \ii_2\times \underbrace{\ii_3\times\cdots\times\ii_3}_{l-2}\times\ii_2$,
with $\times$ denoting the tensor contraction. 
Using $\alpha_i$ to denote the inner bond shared by consecutive tensors, each entry of $\mathcal A$ can be evaluated as
\begin{align}
    \mathcal A(i_1,i_2,\cdots,i_l) &= \sum_{\alpha_1,\alpha_2,\cdots,\alpha_{l-1}}\ii_2(i_1,\alpha_1)\times \ii_3(\alpha_1,i_2,\alpha_2)\times \ii_3(\alpha_2,i_3,\alpha_3)\times\cdots\times\ii_3(\alpha_{l-2},i_{l-1},\alpha_{l-1})\times\ii_2(\alpha_{l-1},i_{l}),\nonumber\\
    & = \sum_{\alpha_1,\alpha_2,\cdots,\alpha_{l-1}}[\delta(i_1,\alpha_1)]\times [\delta(\alpha_1,i_2)\delta(i_2,\alpha_2)]\times [\delta(\alpha_2,i_3)\delta(i_3,\alpha_3)]\times\cdots\times[\delta(\alpha_{l-2},i_{l-1})\delta(i_{l-1},\alpha_{l-1})]\times[\delta(\alpha_{l-1},i_{l})],\nonumber\\
    & = \sum_{\alpha_2,\alpha_3,\cdots,\alpha_{l-1}}[\delta(i_1,i_2)\delta(i_2,\alpha_2)]\times [\delta(\alpha_2,i_3)\delta(i_3,\alpha_3)]\times\cdots\times[\delta(\alpha_{l-2},i_{l-1})\delta(i_{l-1},\alpha_{l-1})]\times[\delta(\alpha_{l-1},i_{l})],\nonumber\\
    & = \sum_{\alpha_3,\alpha_4\cdots,\alpha_{l-1}}[\delta(i_1,i_2)\delta(i_2,i_3)\delta(i_3,\alpha_3)]\times\cdots\times[\delta(\alpha_{l-2},i_{l-1})\delta(i_{l-1},\alpha_{l-1})]\times[\delta(\alpha_{l-1},i_{l})],\nonumber\\
    &=\cdots\nonumber\\
    & = \prod_{a=1}^{l-1}\delta(i_a,i_{a+1}),\nonumber\\
    & = \ii_l(i_1,i_2,\cdots,i_l) .
\end{align}
Applying the above equality to the copy tensor obtained for hidden neurons, one would have $\raisebox{-1ex}{\includegraphics[scale=0.25, trim={0cm 0.2cm 0.2cm 0cm}, clip]{figures/mps1.png}}$; applying it to the copy tensor obtained for visible neurons, one obtains $\raisebox{-1ex}{\includegraphics[scale=0.25, trim={0cm 0.cm 0.cm 0cm}, clip]{figures/mps2.png}}$, as shown in Fig.2(a) and Fig.2(b) in the main text.

\section{Mathematical formulations of the two-dimensional tensor network mapping}
We first give precise values of tensor elements in the obtained two-dimensional tensor network. Suppose the coupling connecting variable $i$ and $j$ is $J_{ij}$. Without loss of generality we set inverse temperature $\beta=1$, and the Boltzmann matrix for variable $i$ and $j$ is 
\begin{align}
\mathbf{B}_{ij}=\left(               
  \begin{array}{cc}   
    e^{ J_{ij}} &e^{- J_{ij}} \\  
    e^{- J_{ij}}& e^{ J_{ij}} \\  
  \end{array}
\right). 
\end{align}
Then the $4$-way tensors which compose the two-dimensional tensor network are obtained by contracting the Boltzmann matrix $\mathbf B$ with two three-way tensors $\ii_3$, as $\raisebox{-1ex}{\includegraphics[scale=0.25, trim={0cm 0.cm 0.cm 0cm}, clip]{figures/element.png}}$, where the brown matrix denotes the Boltzmann matrix $\mathbf B$, and identity matrices are in different colors.
Thus the contraction results $\widehat{\mathbf B}$ can be formulated as
\begin{equation}\widehat{\mathbf B}(a,c,b,d)=\mathbf B(x,y) \times \ii_3(a,x,b)\times \ii_3(c,y,d),\end{equation} 
where the indices $a,b$ come from the hidden neuron and $c,d$ come from the visible neuron.
We first observe that two $\ii_3$ tensors, as well as two sets of indices are completely symmetric, i.e. they are exchangeable. By pinning two indices in one direction, we actually fix one row in the Boltzmann matrix $\mathbf B$. So the resulting matrix should be a diagonal matrix constructed from one row of $\mathbf B$.  
\begin{align} 
\widehat{\mathcal{B}}_{j}^{(i)}(1,:,1,:)=\widehat{\mathcal{B}}_{j}^{(i)}(:,1,:,1)=\widehat{\mathcal{B}}_{j}^{(1)}(1,:,:)=\widehat{\mathcal{B}}_{j}^{(n)}(:,:,1)=\begin{pmatrix}e^{J_{ij}}&0\\0&e^{-J_{ij}}\end{pmatrix}\\ \widehat{\mathcal{B}}_{j}^{(i)}(2,:,2,:)=\widehat{\mathcal{B}}_{j}^{(i)}(:,2,:,2)=\widehat{\mathcal{B}}_{j}^{(1)}(2,:,:)=\widehat{\mathcal{B}}_{j}^{(n)}(:,:,2)=\begin{pmatrix}e^{-J_{ij}}&0\\0&e^{J_{ij}}\end{pmatrix}
\end{align}

There is another way to formula mathematically the mapping process and obtain the same results, which first contracts the Boltzmann matrix to one of the copy tensors, then to the other copy tensor.

First consider the copy tensor $I_n$ associated with an hidden variable $j$ together with associated Boltzmann matrix to an MPS, $$\widetilde {\mathcal H}_j=\{\ii_n,\mathbf B_{1,j},\mathbf B_{2,j},...,\mathbf B_{n,j}   \}.$$ This set of tensors form a classic tensor network format, the canonical polyadic (CP) format~\cite{chi2012tensors}, with CP rank equals $2$. It is known in mathematics that there exists exact conversion from the CP tensor with rank $2$ to an matrix product state with bond dimension $2$ ~\cite{oseledets2011tensor}, which allows us to convert $\widetilde{H}_j$ into 
$$\mathcal H_j\in\mathbb R^{\overbrace{2\times 2\times\cdots\times 2}^{n}}=\begin{pmatrix}e^{J_{1j}}&e^{-J_{1j}}\\e^{-J_{1j}}&e^{J_{1j}}\end{pmatrix}\times \mathcal B^{(2)}_j\times \mathcal B^{(3)}_j\cdots\times \mathcal B^{(n-1)}_j\times \begin{pmatrix}e^{J_{nj}}&e^{-J_{nj}}\\e^{-J_{nj}}&e^{J_{nj}}\end{pmatrix},$$
  where $\mathcal B_j^{(i)}\in\mathbb R^{2\times 2\times 2}$ is a $3$-way tensor with the second leg being the physical leg connecting to $\mathcal V_i$. And $$B[:,1,:]=\begin{pmatrix}e^{J_{nj}}&0\\0&e^{-J_{nj}}\end{pmatrix},\,\,\,\,\,\,\,\,\,\,\,\,B[:,2,:]=\begin{pmatrix}e^{-J_{nj}}&0\\0&e^{J_{nj}}\end{pmatrix}.$$
  For a visible neurons $i$, the conversion to an MPS is written as
  \begin{align}
  \mathcal V_i\in\mathbb R^{\overbrace{2\times 2\times\cdots\times 2}^{m+1}}&=\ii_2\times \overbrace{\ii_3\times \ii_3\times\cdots\times\ii_3}^{m-1}\times \ii_2.
  \end{align}
Then consider contracting $\mathcal H_j$ to $3$-way tensors belonging to every $\mathcal V_i$ with $i=1,...,n$. That is, contract every $3$-way tensor $\mathcal B_j^{(i)}$ and matrices in the boundaries in $\mathcal H_j$ to the corresponding $\ii_3$ in $\mathcal V_i$, resulting to
  $$\widehat {\mathcal H}_j\in\mathbb R^{\overbrace{2\times 2\times\cdots\times 2}^{2n}}_{\underbrace{2\times 2\times\cdots\times 2}_{2n}}=\widehat{\mathcal B}^{(1)}_j\times \widehat{\mathcal B}^{(2)}_j\cdots\times \widehat{\mathcal B}^{(n)}_j,$$ 
  where $\widehat{\mathcal B}_j^{(i)}\in\mathbb R^{2\times 2\times 2 \times 2}$ has the first and third legs connecting tensors $\widehat{\mathcal{B}}_j^{(i-1)}$ and $\widehat{\mathcal{B}}_j^{(i+1)}$, and the second and forth legs connecting to tensors $\widehat{\mathcal{B}}_{j-1}^{(i)}$ and $\widehat{\mathcal{B}}_{j+1}^{(i)}$.  
  With 
  $$\widehat{\mathcal{B}}_{j}^{(i)}[1,:,1,:]=\begin{pmatrix}e^{J_{ij}}&0\\0&e^{-J_{ij}}\end{pmatrix},\,\,\,\,\,\,\,\,\,\,\,\,\widehat{\mathcal{B}}_{j}^{(i)}[2,:,2,:]=\begin{pmatrix}e^{-J_{ij}}&0\\0&e^{J_{ij}}\end{pmatrix}.$$
  On the left boundary, tensors are three way tensors $\widehat{\mathcal B}_j^{(1)}\in\mathbb R^{ 2\times 2 \times 2}$ with the first leg and the last leg connecting $\widehat{\mathcal B}_{j-1}^{(1)}$ and $\widehat{\mathcal B}_{j+1}^{(1)}$  respectively, with $${\mathcal{B}}_{j}^{(i)}[:,1,:]=\begin{pmatrix}e^{J_{1j}}&0\\0&e^{-J_{1j}}\end{pmatrix},\,\,\,\,\,\,\,\,\,\,\,\,\widehat{\mathcal{B}}_{j}^{(i)}[:,2,:]=\begin{pmatrix}e^{J_{1j}}&0\\0&e^{-J_{1j}}\end{pmatrix}.$$
  On the right boundary, tensors are three way tensors $\widehat{\mathcal B}_j^{(n)}\in\mathbb R^{ 2\times 2 \times 2}$ with the second leg and the last leg connecting $\widehat{\mathcal B}_{j-1}^{(n)}$ and $\widehat{\mathcal B}_{j+1}^{(n)}$  respectively, with $${\mathcal{B}}_{j}^{(n)}[1,:,:]=\begin{pmatrix}e^{J_{nj}}&0\\0&e^{-J_{nj}}\end{pmatrix},\,\,\,\,\,\,\,\,\,\,\,\,\widehat{\mathcal{B}}_{j}^{(n)}[2,:,:]=\begin{pmatrix}e^{J_{nj}}&0\\0&e^{-J_{nj}}\end{pmatrix}.$$

\section{Two dimensional tensor network representation of the RBM with external fields}
\begin{figure}[htb]
\centering
\includegraphics[width=\columnwidth]{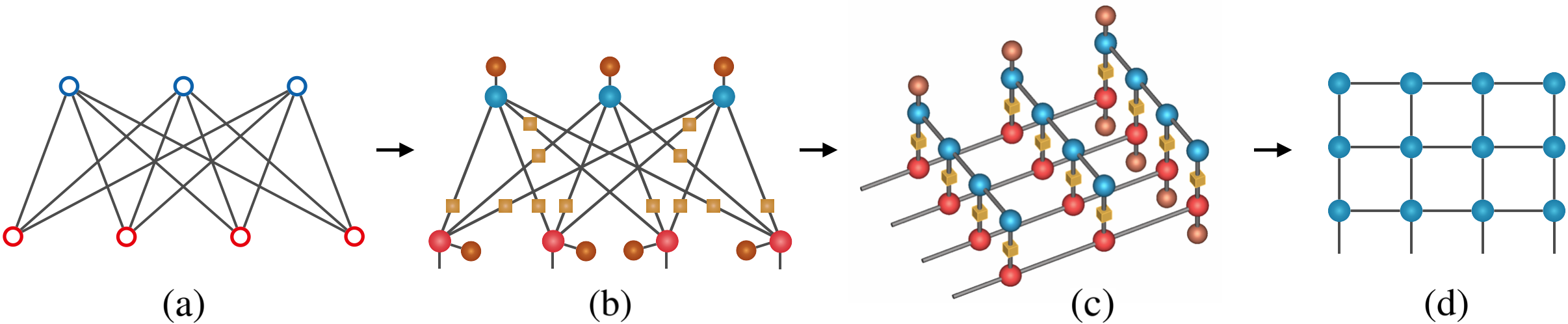}
\caption{Illustration of the mapping from an RBM with external fields to a two-dimensional tensor network. (a) The graphical representation of an RBM with four visible neurons (red circles) and three hidden neurons (blue circles). (b) The tensor network representation of the RBM with external fields. Red and blue nodes are copy tensors corresponding to visible and hidden variables, respectively; dark brown nodes represent their external field vectors; brown squares denote Boltzmann matrices between visible and hidden nodes. (c) The three-dimensional diagram after converting each copy tensor to an MPS.  (d) The final two-dimensional tensor network representation of the RBM with external fields.}
\label{fig:rbm_bias}
\end{figure}

\begin{figure}[htb]
\centering
\subfigure[]{
\label{fig:bias_mps1}
\includegraphics[width=0.3\columnwidth]{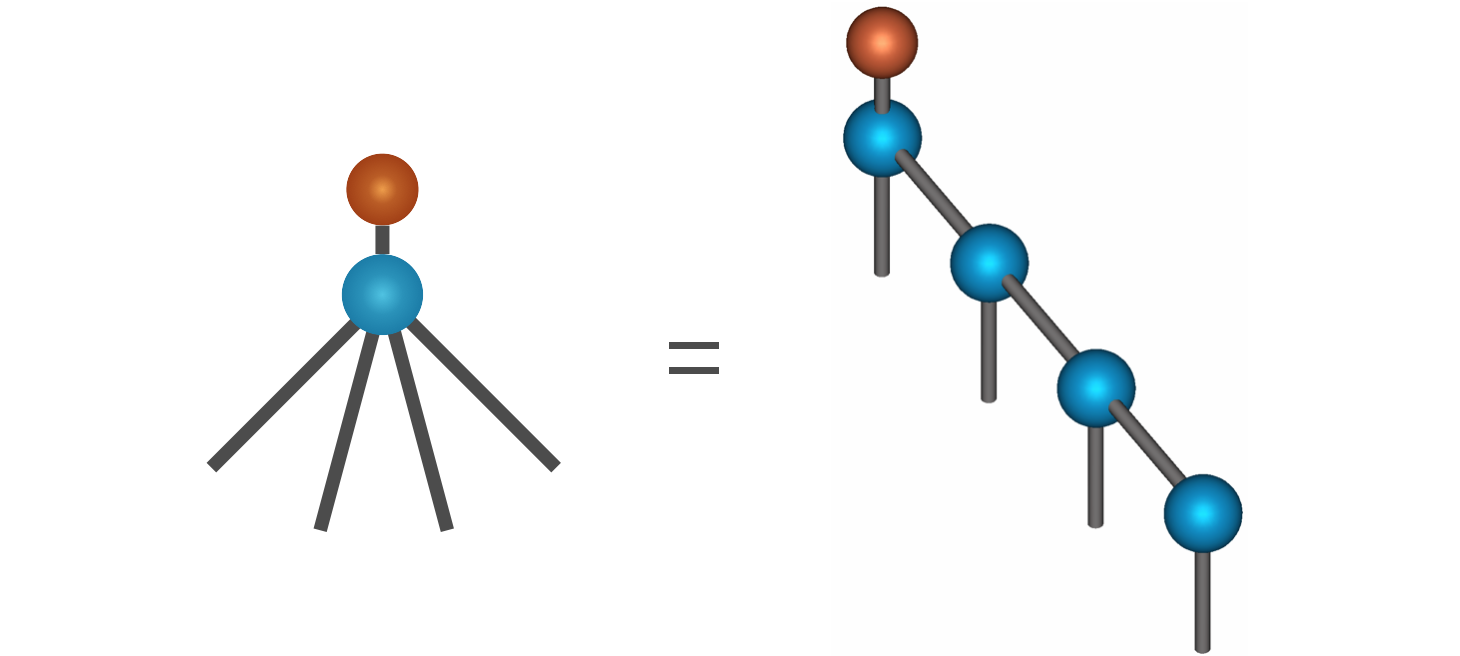}}
\quad
\subfigure[]{
\label{fig:bias_mps2}
\includegraphics[width=0.3\columnwidth]{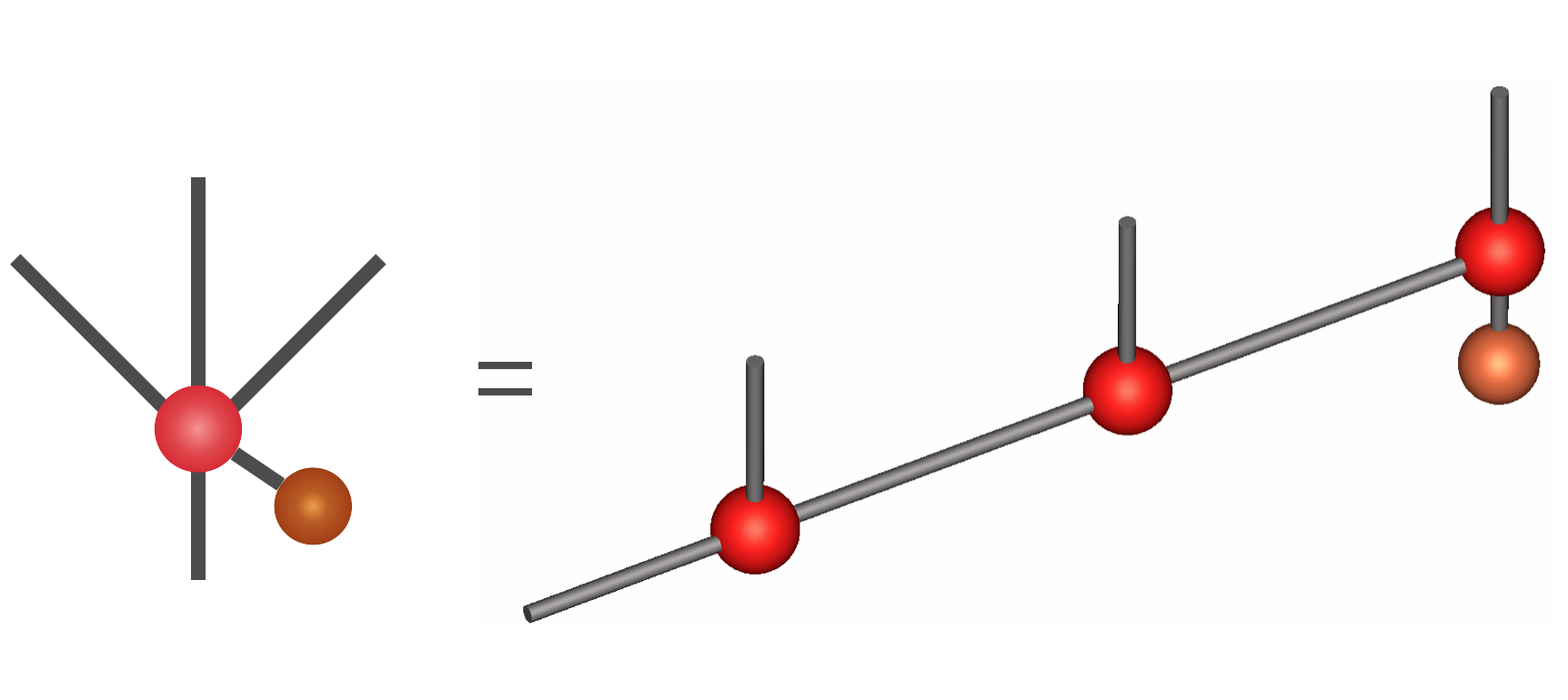}}
\quad
\subfigure[]{
\label{fig:bias_element}
\includegraphics[width=0.3\columnwidth]{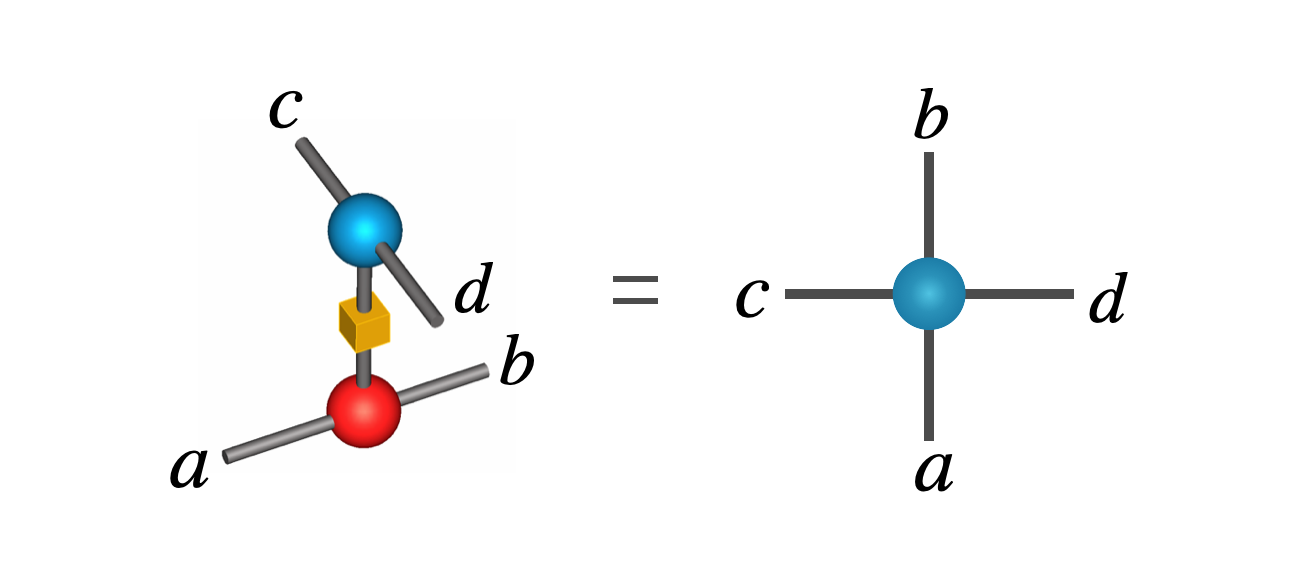}}
\caption{(a) and (b) are schematic diagrams of converting a copy tensor with an external field vector to an MPS, (a) for hidden variables and (b) for visible variables. (c) Converting the three-dimensional structure to a single tensor.}
\label{fig:bias_mps}
\end{figure}

More generally, in RBM and DBM there is an associated external field (bias) for each visible neuron and hidden neuron. Here we discuss the conversion from RBM with external fields (bias) to two-dimensional tensor network, following the same procedure in the main text, as pictorially represented in Fig.~\ref{fig:rbm_bias}.
First, the RBM (with the graphical representation in Fig.~\ref{fig:rbm_bias}(a)) is converted to a tensor network representation (Fig.~\ref{fig:rbm_bias}(b)) with the same topology. The blue copy tensors come from visible neurons, and the red copy tensors come from hidden variables, each of them is connected to an external field denoted by a brown tensor (which actually denotes a vector, with one index). To be more precise, the external field vector of node $i$ is $(\theta_i,-\theta_i)$ and $\theta_i$ is the external field of neuron $i$. Then, each copy tensor is converted to an MPS using the same method described in the main text, as shown in Fig.~\ref{fig:bias_mps}(a) and (b). Since it does not matter which tensor of the MPS we put the external field vector on, without loss of generality we put them on the first tensor of the MPS. Fig.~\ref{fig:rbm_bias}(c) represents the tensor network after converting copy tensors to MPSs. Finally, by contracting the Boltzmann matrices with connected copy tensors, we obtain a two-dimensional tensor network with size $m\times n$, where the number of rows $m$ equal the number of hidden variables, and the number of columns $n$ equals the number of visible variables.
\section{Mapping DBMs with many hidden layers  to two-dimensional tensor networks}
\begin{figure}[htb]
\centering
\includegraphics[width=\columnwidth]{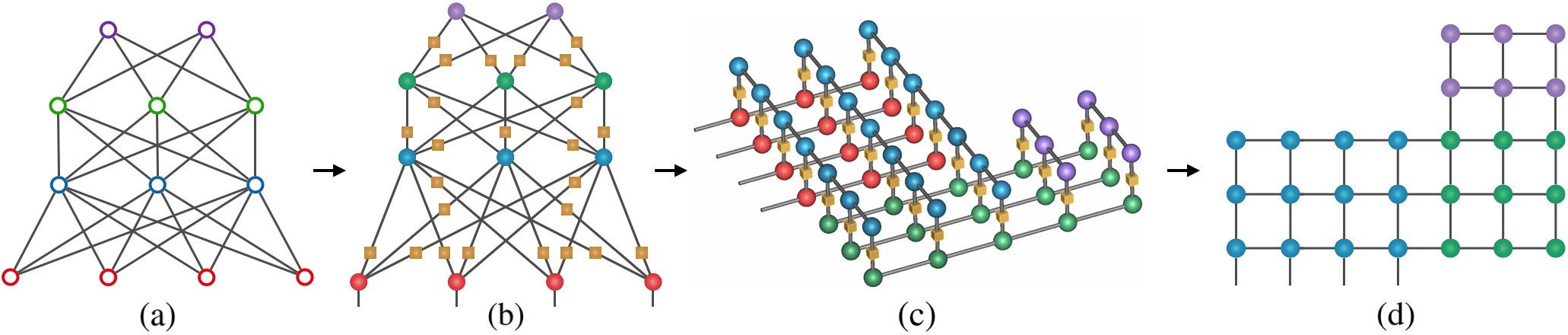}
\caption{Illustration of mapping a DBM with three hidden layers to a two-dimensional tensor network. (a) The graphical representation of a DBM with three hidden layers. Red circles are visible variables. Blue, green, and purple circles are hidden variables. (b) The tensor representation of DBM. All nodes are copy tensors, and brown squares denote Boltzmann matrices between variables of two adjacent layers. Converting each copy tensor to an MPS, we obtain (c), where the color of the MPS corresponds to the color of the tensor in (b) one-to-one. Compress the tensor network of (c) in the vertical direction, and then arrive at figure (d), a two-dimensional tensor network representation. Blue tensors indicate that they are from the first hidden layer, green tensors denote that they are from the second hidden layer, and purple tensors denote that they are from the third hidden layer.}
\label{fig:dbm2_all}
\end{figure}

Similar to the conversion procedure of DBM with two hidden layers, there are three steps to map a DBM with three hidden layers to a two-dimensional tensor network. As shown in Fig.~\ref{fig:dbm2_all}(a), consider a DBM with one visible layer and three hidden layers. Replacing each variable with a copy tensor and putting a Boltzmann matrix on each edge, we obtain the tensor representation (Fig.~\ref{fig:dbm2_all}(b)) of the DBM with the same topology. Then, converting copy tensors to MPSs, we arrive at a three-dimensional tensor network (Fig.~\ref{fig:dbm2_all}(c)). Notice that the length of the MPS is equal to the order of the original copy tensor. Finally, by contracting the three dimensional tensor network in the vertical direction, we obtain a two-dimensional tensor network, as shown in Fig.~\ref{fig:dbm2_all}(d) where we have used the same colors for tensors as the variables in the DBM.
\par

\begin{figure}[htb]
\centering
\includegraphics[width=0.9\columnwidth]{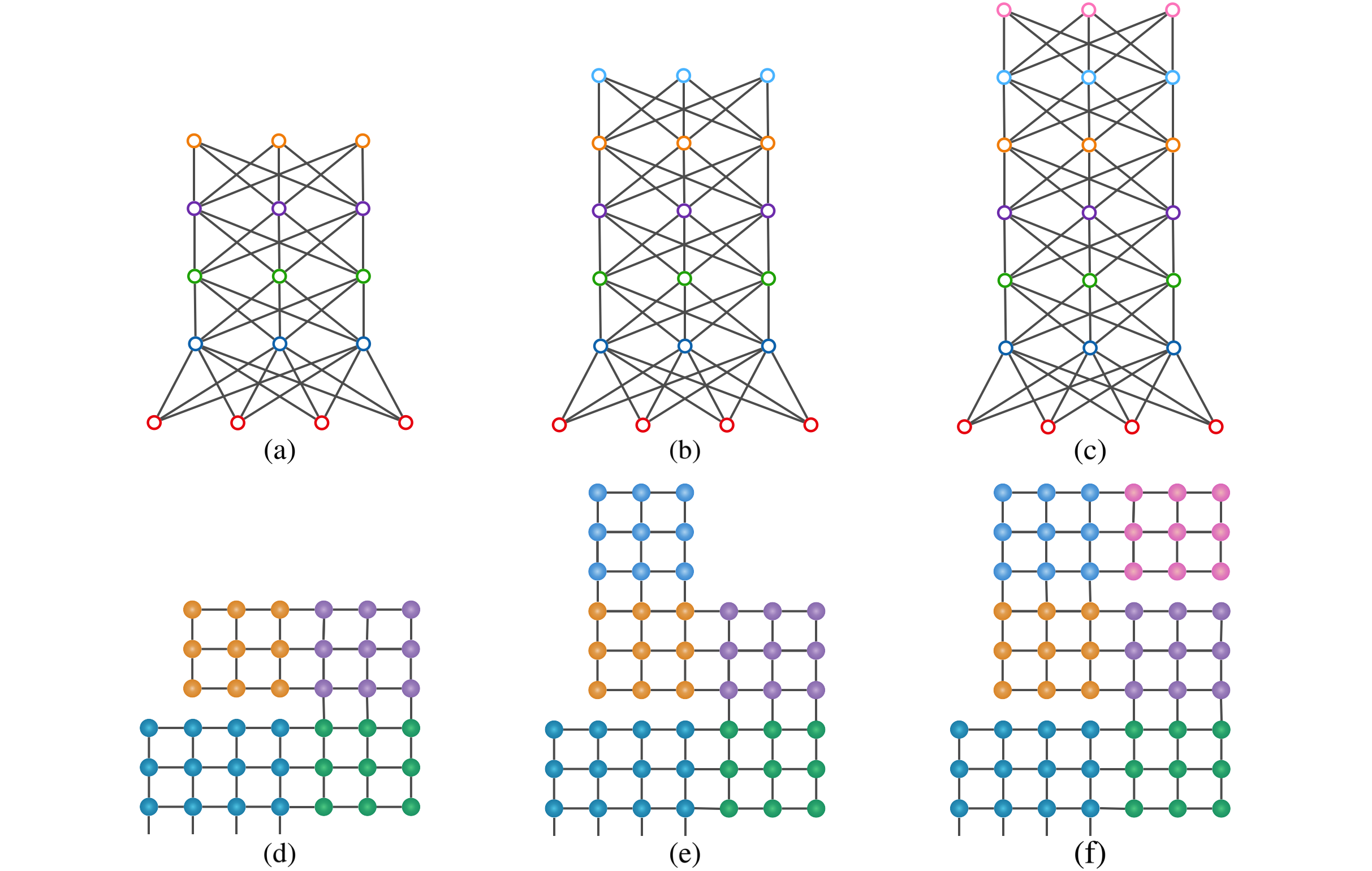}
\caption{(a) A DBM with one visible layer (red circles) and four hidden layers (blue, green, purple and orange circles). (b) A DBM with five hidden layers. (c) A DBM with six hidden layers. (d),(e),(f) are two-dimensional tensor network representations of the DBM in (a), (b) and (c) respectively.}
\label{fig:dbm_more}
\end{figure}

We can map a DBM with an arbitrary number of hidden layers to a two-dimensional tensor network following the same procedure introduced above, resulting in a tensor networks with a shape depending on the topology of the DBM. In Fig.~\ref{fig:dbm_more}, we illustrate the conversions from five-layer, six-layer, and seven-layer DBMs. In the figure, (a)-(c) are graphical representations of DBMs, and (d)-(f) are their two-dimensional tensor network representations respectively. As the number of hidden layers increases, we can make the two-dimensional tensor network gradually stacks upwards in an ``S'' shape.
\section{Contracting two-dimensional tensor networks}
There are many methods for contracting two-dimensional tensor networks. We chose the boundary matrix product state approach (BMPS). With limited computing resources, it can reasonably control the number of errors made in the cut-off approximations through the canonical form~\cite{schollwock2011density}. The key component of the BMPS method is identifying a boundary of the two dimensional tensor network, treating the tensors on the boundary as a matrix product state, and tensors inside the tensor network as a stack of matrix product operators (MPO). Then the MPOs are contracted to the MPS on the boundary once a time, resulting in an MPS with larger bound dimensions, until finally arriving at a one-dimensional tensor network (i.e. an MPS without open indices), which can be contracted to a scalar straightforwardly. When the MPS contains bound dimensions that are larger than the threshold $\dmax$, we use singular value decompositions to reduce the bound dimension.

\begin{figure}[htb]
\centering
\includegraphics[width=\columnwidth]{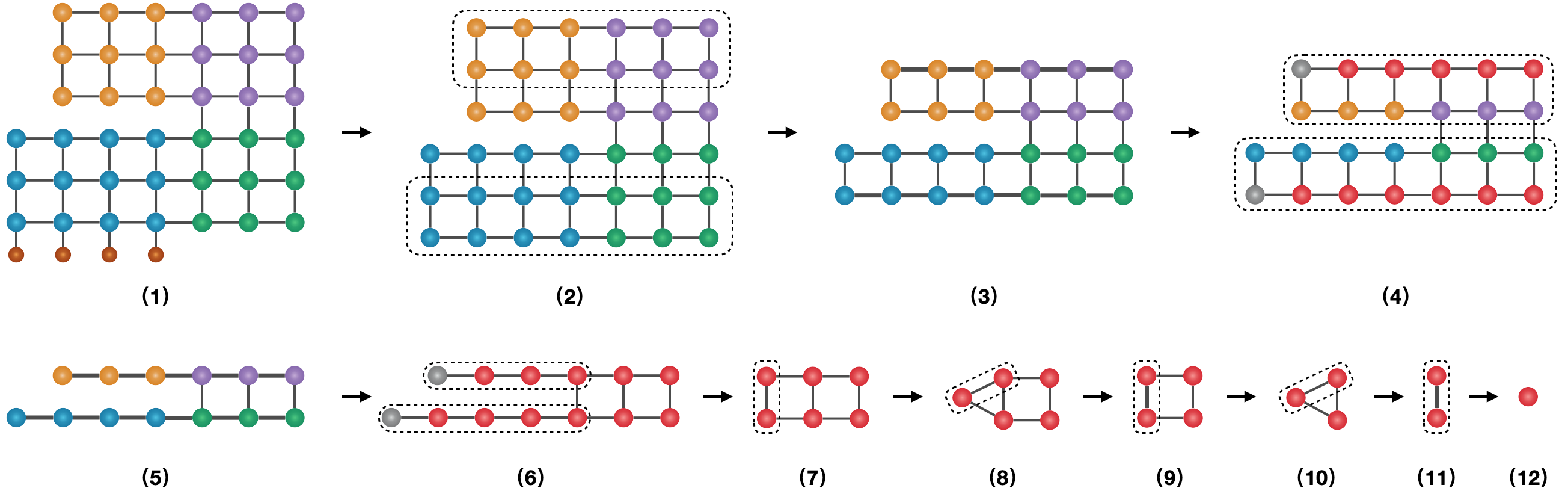}
\caption{Illustration of the contraction process from a two-dimensional tensor network (mapped from a DBM with four hidden layers) to a scalar representing the partition function of the DBM, using the BMPS method.}
\label{fig:bmpsZ}
\end{figure}

\begin{figure}[htb]
\centering
\includegraphics[width=0.95\columnwidth]{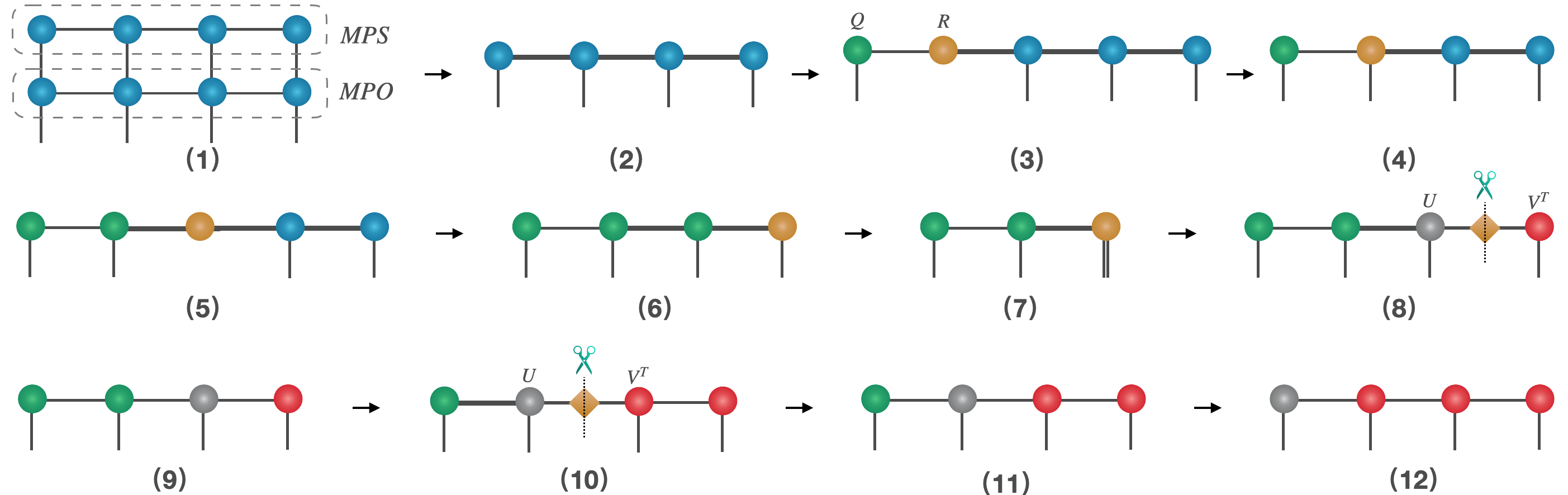}
\caption{Detailed process of $\emph{merging}$, $\emph{canoicalization}$ and $\emph{truncation}$. $\emph{Merging}$ consists of (1) and (2); $\emph{canoicalization}$ refers to (3)-(6); $\emph{truncation}$ means (7)-(12).}
\label{fig:bmps}
\end{figure}

Taking a DBM with four hidden layers as an example, we apply the BMPS method to estimate the partition function defined on the model. Fig.~\ref{fig:dbm_more}(d) displays a two-dimensional tensor network representation of the unnormalized probability $\wttp$. As we have explained in the main text, the partition function is obtained by the contraction of $\wttp$ with $n$ $(1,1)$ vectors. As shown in Fig.~\ref{fig:bmpsZ}(1), dark brown nodes are $(1,1)$ vectors, and contracting with these vectors, we get the two-dimensional tensor network form of the partition function (Fig.~\ref{fig:bmpsZ}(2)). An efficient way to contract the network is to divide it into two parts, and the operations applied to each part are the same. The key operations of contraction are introduced in Fig.~\ref{fig:bmps}, which consists of $\emph{merging}$ (Fig.~\ref{fig:bmps}(1),(2)), $\emph{canonicalization}$ (Fig.~\ref{fig:bmps}(3)-(6)) and $\emph{truncation}$ (Fig.~\ref{fig:bmps}(7)-(12)). \par
From Fig.~\ref{fig:bmps}(1), we can see that there is an MPS (with bound dimension $D$) on the boundary, stacked on top of an MPO. Contracting the MPS with the MPO, we obtain a new MPS (Fig.~\ref{fig:bmps}(2)) with bond dimension $D^2$ . In practice, we need to set an upper limit $\dmax$ to control the computational cost. If $D^2$ is larger than $\dmax$, we perform cut off on bond dimension through singular value decomposition (SVD) operations. Fig.~\ref{fig:bmps}(3)-(12) describe the details of this step, where from (3) to (6) the MPS is converted to a canonical form; from (7) to (12) the bond dimension is reduced using singular value decompositions. It is worth noting that the canonical form of the MPS is critical for reducing truncation errors. On the one hand, the canonical form makes the computation of expectations and correlations involving only local terms, and on the other hand, it makes the truncation operations in the global scope of the whole MPS rather than two local tensors. 
In this work, the canonical form is achieved by applying QR decompositions. As shown in Fig.~\ref{fig:bmps}(3) and (4), QR decomposition is applied to the left-most tensor, then absorbs the $R$ matrix (brow node) to the right tensor (blue node). Repeating the decomposition and absorption operations for each tensor to the right-most one, we arrive at Fig.~\ref{fig:bmps}(6), where all the tensors except the right-most one, are isometries. 
Next, starting from the right-most tensor, we apply SVD to truncate the bond dimension. We first merge the two right-most tensors to produce a new tensor $\mathcal{T}$ (Fig.~\ref{fig:bmps}(7)), then perform SVD on it (Fig.~\ref{fig:bmps}(8)), 
\begin{align}
\mathcal{T}=USV^T,\quad S=diag(\lambda)
\end{align}
where $\lambda$ is the set of sorted singular values. If the number of the eigenvalues is larger than $D^2$, we only keep the first $D^2$ values and throw away the rest. Correspondingly, $U$ matrix keeps $D^2$ columns, and  $V^T$ matrix keeps $D^2$ rows. The new tensors are
\begin{align}
    \mathcal{T}_1=\widetilde{U}\sqrt{\widetilde{S}},\quad\mathcal{T}_2=\sqrt{\widetilde{S}}\widetilde{V^T}.
\end{align}
where $\widetilde{U}, \widetilde{S}, and \,\widetilde{V}^T$ label the truncated $U, S, and \,V^T$. Repeating the steps (7)-(9) to the left-most tensor, we obtain the truncated MPS Fig.~\ref{fig:bmps}(12).\par
The tensor network shown in Fig.~\ref{fig:bmpsZ}(2) is formed by stacking MPSs and MPOs. We focus on two boundaries which are on the top and on the bottom. Starting from these two MPSs, and contracting them with the connected MPOs (\emph{merging}), we obtain two new MPSs shown in Fig.~\ref{fig:bmpsZ}(3). Then, performing \emph{canonicalization} and \emph{truncation} on the new MPSs, we get Fig.~\ref{fig:bmpsZ}(4). Repeating Fig.~\ref{fig:bmpsZ}(2)-(4) steps, the tensor network evaluates to Fig.~\ref{fig:bmpsZ}(6). By contracting tensors according to the dashed boxes in Fig.~\ref{fig:bmpsZ}(6)-(11), we get the final result, a scalar (Fig.~\ref{fig:bmpsZ}(12)) representing the partition function of the original DBM.
\end{document}